\documentclass[sigconf]{acmart}

\usepackage{amsmath,amssymb,amsfonts}
\usepackage{algorithmic}
\usepackage{graphicx}
\usepackage{textcomp}
\usepackage{xcolor}
\usepackage{booktabs}
\usepackage{colortbl}
\usepackage{multirow}
\usepackage{bm}
\usepackage[justification=centering]{caption}
\usepackage{balance}
\usepackage{float}
\usepackage{comment}

\usepackage{pgfplots}

\AtBeginDocument{%
  \providecommand\BibTeX{{%
    \normalfont B\kern-0.5em{\scshape i\kern-0.25em b}\kern-0.8em\TeX}}}

\copyrightyear{2021}
\acmYear{2021}
\setcopyright{acmcopyright}\acmConference[EASE 2021]{Evaluation and Assessment
in Software Engineering}{June 21--23, 2021}{Trondheim, Norway}
\acmBooktitle{Evaluation and Assessment in Software Engineering (EASE 2021), June
21--23, 2021, Trondheim, Norway}
\acmPrice{15.00}
\acmDOI{10.1145/3463274.3463333}
\acmISBN{978-1-4503-9053-8/21/06}

\begin{document}

\title{Self-Claimed Assumptions in Deep Learning Frameworks: An Exploratory Study}

\author{Chen Yang$^{1,2}$, Peng Liang$^{1*}$, Liming Fu$^{1}$, Zengyang Li$^{3}$}
\affiliation{%
  \institution{$^{1}$School of Computer Science, Wuhan University, Wuhan, China}
  \institution{$^{2}$IBO Technology (Shenzhen) Co., Ltd., Shenzhen, China}
  \institution{$^{3}$School of Computer Science \& Hubei Provincial Key Laboratory of Artificial Intelligence and Smart Learning, Central China Normal University, Wuhan, China}
  \institution{\{cyang, liangp, limingfu\}@whu.edu.cn,  zengyangli@ccnu.edu.cn}
}

\renewcommand{\shortauthors}{C. Yang et al.}




\begin{abstract}
Deep learning (DL) frameworks have been extensively designed, implemented, and used in software projects across many domains. However, due to the lack of knowledge or information, time pressure, complex context, etc., various uncertainties emerge during the development, leading to assumptions made in DL frameworks. Though not all the assumptions are negative to the frameworks, being unaware of certain assumptions can result in critical problems (e.g., system vulnerability and failures, inconsistencies, and increased cost). As the first step of addressing the critical problems, there is a need to explore and understand the assumptions made in DL frameworks. To this end, we conducted an exploratory study to understand self-claimed assumptions (SCAs) about their distribution, classification, and impacts using code comments from nine popular DL framework projects on GitHub.
The results are that: (1) 3,084 SCAs are scattered across 1,775 files in the nine DL frameworks, ranging from 1,460 (TensorFlow) to 8 (Keras) SCAs. (2) There are four types of validity of SCAs: Valid SCA, Invalid SCA, Conditional SCA, and Unknown SCA, and four types of SCAs based on their content: Configuration and Context SCA, Design SCA, Tensor and Variable SCA, and Miscellaneous SCA. (3) Both valid and invalid SCAs may have an impact within a specific scope (e.g., in a function) on the DL frameworks. Certain technical debt is induced when making SCAs. There are source code written and decisions made based on SCAs. 
This is the first study on investigating SCAs in DL frameworks, which helps researchers and practitioners to get a comprehensive understanding on the assumptions made. We also provide the first dataset of SCAs for further research and practice in this area.
\end{abstract}

\keywords{Self-Claimed Assumption, Deep Learning Framework, GitHub}

\maketitle
\section{Introduction} \label{introduction}
Deep learning (DL) is a branch of machine learning, which has been extensively used in various areas as a key component, for example, DL can be used for tasks such as image classification and object detection in computer vision. To improve the efficiency of developing DL based applications, many popular frameworks (e.g., TensorFlow~\cite{Abadi2016} and PyTorch~\cite{Steiner2019}) have been designed, implemented, and used in software projects. However, due to the lack of knowledge or information, time pressure, complex context, etc., various uncertainties emerge during the development, leading to assumptions made in DL frameworks. Though not all the assumptions are negative to the frameworks, being unaware of certain assumptions can result in critical problems, including system vulnerability and failures, inconsistencies, misuse and misunderstanding of the system, and increased cost~\cite{Yangsms2018}. For example, in TensorFlow (version: ``\textit{tensorflow\-gpu==2.0.0}''), the placer algorithm assumes that any tensor can be copied between different devices. However, this assumption is invalid (e.g., for dataset variant tensors). As recorded on the TensorFlow issue tracking system (e.g., Issue \#34112\footnote{https://github.com/tensorflow/tensorflow/issues/34112} and Issue \#34519\footnote{https://github.com/tensorflow/tensorflow/issues/34519}), some users encountered system failures since they were not aware of the assumption. Moreover, Issue \#34112 and \#34519 were first reported on 2019-11-09 and 2019-11-22, respectively. Issue \#34519 was closed on 2020-09-04, taking more than nine months and involving seventeen developers and users and Issue \#34112 is still open (i.e., lasting more than 17 months) with thirteen developers and users involved. As the first step of addressing the critical problems, there is a need to explore and understand the assumptions made in DL frameworks.

Assumptions are subjective in software development~\cite{Yangsms2018}, which is the major reason that stakeholders may have a different understanding of the assumption concept. For example, Roeller \textit{et al.}~\cite{Roeller2006} stated: ``\textit{From one perspective or stakeholder, we may denote something as an assumption, while that same thing may be seen as a design decision from another perspective.}'' Also, assumptions are context dependent~\cite{Yangsms2018}. The same assumption could be valid in one project, and invalid in another project because the context changes; or an assumption in one project is not an assumption in another project. Based on the nature of assumptions, unless they are explicitly claimed (e.g., using phrases like ``\textit{the assumption is}'' or ``\textit{it is assumed that}''), it is difficult to judge whether a piece of information is an assumption in software development~\cite{Yangsms2018}. Therefore, in this paper, we focused on those explicitly claimed assumptions by developers, i.e., self-claimed assumptions (SCAs), and conducted an exploratory study to understand SCAs about their distribution, classification, and impacts using code comments from nine popular DL framework projects on GitHub. Additionally, though there are many related studies of assumptions in software development, very few datasets have been available to the community, and the size of the provided datasets are small (e.g., tens or hundreds of assumptions). This impedes researchers from analyzing assumptions in software development and practitioners from efficiently using and managing assumptions in their projects. Therefore, our exploratory study provides a dataset with 3,084 assumptions for further research and practice in this area (e.g., developing methods for automatic extraction and classification of assumptions). 

In the rest of the paper, Section \ref{relatedwork} summarizes related work on DL frameworks and assumptions in software development. Section \ref{studydesign} describes the study design. The results and discussions are provided in Section \ref{results} and Section \ref{discussion}, respectively. Section \ref{threatstovalidity} analyzes the threats to the validity of the study. Section \ref{conclusions} concludes the study with future directions.

\section{Related Work} \label{relatedwork}
\subsection{Deep Learning Frameworks} \label{DLframworks}
Erickson \textit{et al.}~\cite{Erickson2017} first provided a set of criteria for evaluating a DL framework, including programming language, documentation, development environment, execution speed, training speed, GPU support, maturity level, model library, GitHub commits, and GitHub contributors. Then they briefly evaluated twelve DL frameworks (e.g., Caffe, DL4J, TensorFlow, Theano, and Keras) using the criteria. Parvat \textit{et al.}~\cite{Parvat2017} conducted a survey on Theano, DL4J, Caffe, Torch, and TensorFlow regarding different aspects (e.g., creator, programming language, interface, platform, parallelization, pre-trained model, community, and documentation). Shi \textit{et al.}~\cite{Shi2017} compared Caffe, CNTK, MXNet, TensorFlow, and Torch through experiments regarding their performance on various types of neural networks and hardware platforms. The results show that all the five DL frameworks can efficiently use GPUs in different tasks, while no framework can outperform the others in all aspects. Wu \textit{et al.}~\cite{Wu2018} conducted a comparative analysis and characterization on TensorFlow, Caffe, Torch, and Theano, including factors that have an impact on the runtime performance and accuracy of the frameworks; and relationships between different settings of CPU/GPU usage and batch size of data. Zhang \textit{et al.}~\cite{Zhang2018} conducted an empirical study regarding 175 bugs of TensorFlow collected from Stack Overflow and GitHub. They studied the root causes and symptoms of the bugs; and the solutions for bug detection and localization. Islam \textit{et al.}~\cite{Islam2019} also focused on the bugs of DL frameworks collected from both Stack Overflow and Github, and they studied Caffe, Keras, TensorFlow, Theano, and Torch to understand the types, root cause, and impacts of the bugs; bug-prone stage of DL pipeline; and related anti-patterns in the frameworks. Guo \textit{et al.}~\cite{Guo2018} explored how various DL frameworks support DL-based application development and deployment. They identified practical guidelines, challenges, and future directions of developing and using DL frameworks. Zhang \textit{et al.}~\cite{Zhang2019} focused on challenges of developing DL applications based on TensorFlow, PyTorch, and DL4J. They found that program crashes, model migration, and implementation questions are the most discussed challenges. Nejadgholi and Yang~\cite{Nejadgholi2019} conducted an empirical study on oracle approximations through TensorFlow, Theano, PyTorch, and Keras. Their study covers the prevalence of oracle approximations, diversity of test oracles and thresholds used in oracle approximations, and code changes and the reasons related to oracle approximation. Liu \textit{et al.}~\cite{Liu2020} analyzed the code comments from TensorFlow, Keras, Caffe, PyTorch, MXNet, CNTK, and DL4J on GitHub. They focused on technical debt of the frameworks regarding their prevalence, types, and distribution. Han \textit{et al.}~\cite{Han2020} mined the topics of TensorFlow, PyTorch, and Theano on Stack Overflow and GitHub. They compared the topics across different frameworks and platforms. Ren \textit{et al.}~\cite{Ren2020} conducted an exploratory study of the issues regarding performance, memory, and resource usage in ten machine learning (including DL) frameworks.
\textbf{In the related work above, TensorFlow, Theano, PyTorch, Caffe, MXNet, Keras, CNTK,  DL4J, and PaddlePaddle are the most popular DL frameworks, which formed the basis of the DL frameworks explored in this study.}
\subsection{Assumptions in Software Development}
We~\cite{Yangsms2018} conducted a systematic mapping study on assumptions and their management in software development, including understanding of assumptions, assumption management activities, approaches and tools, stakeholders, benefits, challenges, and lessons learned of assumption management, and consequences caused by not well-managed assumptions~\cite{Yangsms2018}. We also conducted an industrial survey regarding the integration of agile practices into software architectural assumption management~\cite{Yang2019}. Bazarhanova and Smolander~\cite{Bazarhanova2020} performed a review on assumptions, while their scope was limited to non-technical assumptions (i.e., interplay between technology and business, inter-organizational aspects, and governance) in digital identity management systems. Xiong \textit{et al.}~\cite{Xiong2018} conducted an exploratory study on assumptions in open source software development. They identified 832 assumptions from the developer mailing list of a popular OSS project named Hibernate, and they further analyzed assumption expression, classification, trend of assumptions over time, and related software artifacts. Landuyt and Joosen~\cite{Landuyt2020} conducted a descriptive study on assumptions made during the application of a threat modeling framework (i.e., LINDDUN), which allows to identify privacy-related design flaws in software architecture design.
For assumption management in software development, most of the related work focuses on Assumption Making, Documentation, and Evaluation~\cite{Yangsms2018}. Fu \textit{et al.}~\cite{Fu2020} presented a tool named UACFinder to automatically identify potential syntactic carriers (e.g., constant variables and frequently read variables) of unspecified assumptions in software design models. Fassi \textit{et al.}~\cite{Fassi2020} proposed an assumption network-based approach for allocating and managing margins in system design. The approach is composed of five activities, i.e., Assumption Identification, Assumption Analysis, Assumption Dependency Capturing (including an assumption network), Margin Allocation and Management, and Assumption Network Maintenance. Gaaloul \textit{et al.}~\cite{Gaaloul2020} proposed an approach (i.e., EPIcuRus) to automatically analyze environment assumptions for software components. The approach employs search-based testing, machine learning, and model checking techniques. Jeong \textit{et al.}~\cite{Jeong2020} also focused on environment assumptions, and they proposed a log-based testing approach to identify faults caused by invalid environment assumptions in model-based development. Li \textit{et al.}~\cite{Li2019} conducted an experiment to evaluate the performance of seven machine learning methods (e.g., Support Vector Machine, Logistic Regression, and Perceptron) in automatically classifying software development assumptions. We~\cite{Yangprocess2018} proposed a general process for software architectural assumption management, and evaluated the process with two case studies regarding the ease of understanding and the effort of conducting the process, as well as the effectiveness of using the process to make architectural assumptions explicitly and to identify and reduce invalid architectural assumptions.
There are also studies (e.g.,~\cite{Jong2019}, ~\cite{Rizkiyanto2016}, and ~\cite{Tang2018}) that treat assumptions as second-class entities in software development, for example, a kind of rationale for software design decisions, and include assumption management as part of a single step in software development.
\textbf{Assumptions and their management are important in software development. Not well-managed assumptions may lead to various problems. To the best of our knowledge, there is no existing work that focuses on assumptions made in DL frameworks. This work is the first step to identify, classify, and analyze assumptions in the context of DL framework projects.}

\section{Study Design} \label{studydesign}
\subsection{Objective and Research Questions}
We followed the guidelines proposed by Runeson \textit{et al.}~\cite{Runeson2012} to design this study.
The objective of the case study is to analyze code comments for the purpose of exploration with respect to SCAs and their distribution, classification, and impacts from the point of view of software developers in the context of nine popular DL framework projects on GitHub. The research questions (RQs) are formulated according to the objective of this study:

\textbf{RQ1: What is the distribution of SCAs in deep learning frameworks?} According to the subjective and context dependent characteristics of assumptions in software development, the distribution of assumptions made may vary greatly in different projects. The results of this RQ helps to understand the prevalence of SCAs in DL frameworks.

\textbf{RQ2: What are the types of SCAs in deep learning frameworks?} Developers constantly make assumptions in their work, and different types of assumptions may have different impacts on software development. This RQ intends to provide a basic understanding regarding the categories of SCAs in DL frameworks.

\textbf{RQ3: What are the impacts of SCAs in deep learning frameworks?} Assumptions are related to various types of software artifacts and not well-managed assumptions may have negative impacts on software development. This RQ helps to explore such impacts of SCAs in DL frameworks.

\subsection{Data Extraction and Analysis}
The study focused on DL frameworks instead of DL applications built on DL frameworks. We selected nine of the most popular DL frameworks (i.e., TensorFlow, Theano, PyTorch, Caffe, MXNet, Keras, CNTK, DL4J, and PaddlePaddle) as the cases to be studied based on the literature (see Section \ref{DLframworks}).

The extracted data items and their description regarding each DL framework and each SCA are provided in Table \ref{tab_dataitem_dl} and Table \ref{tab_dataitem_sca}, respectively. The data extraction steps are the following.

(1) Downloaded the latest version of source code from the DL framework projects on GitHub. Extracted the following data items: ``\textit{Framework}'', ``\textit{Release}'', and ``\textit{Languages}''.

(2) Used Visual Studio to collect ``\textit{Number of Files}'' and Tokei
to collect ``\textit{Lines of Comments}'' and ``\textit{Lines of Code}''.

(3) Since the focus of this study is SCAs, we used the following search terms stemmed from \textit{assumption} to search SCAs from the downloaded files in each DL framework project: (a) \textit{assumption}; (b) \textit{assumptions}; (c) \textit{assume}; (d) \textit{assuming}; (e) \textit{assumed}; and (f) \textit{assumes}. Step (1), (2), and (3) were conducted by the third author and the first author checked the results.

(4) We conducted a two-round pilot data extraction on PyTorch (i.e., the first round with 40 search results) and TensorFlow (the second round with 40 search results), respectively. In both rounds, the first and third authors independently extracted data from the search results. The extracted data was then reviewed by the other authors. All the conflicts were resolved through discussions. An example of the conflicts is whether SCAs from .td files should be extracted. Since .td files are a type of documentation that can be used to generate source code through tablegen tools, we only included the generated source code but excluded the .td files.
In the first round of the pilot data extraction, the inter-rater agreement on whether to include a piece of data that was measured using the Cohen's Kappa coefficient~\cite{Cohen1960} is 0.750. We (a) refined the data items for data extraction (e.g., we added a new data item: ``\textit{Framework}'' and separated two data items ``\textit{Owner}'' and ``\textit{Constraint}'' from the data item ``\textit{Assumption}''), (b) reached an agreement on the understanding of the data items and data extraction criteria, and (c) reduced inconsistencies of data extraction among the authors. 
In the second round of the pilot data extraction, the inter-rater agreement on whether to include a piece of data that was measured using the Cohen's Kappa coefficient~\cite{Cohen1960} is 0.950, which indicates an almost perfect agreement between the first and third authors. The reasons are: we used the search terms stemmed from \textit{assumption} and the search results were either specific names (e.g., variable names) or SCAs; and it was easy to judge whether a search result is from code comments or other sources since we reached an agreement on the data extraction criteria: (a) Exclude assumptions from other sources, such as source code (e.g., a variable named ``\textit{assumed}'' in Caffe), log (e.g., ``\textit{logging.warning(``CrossShardOptimizer should be used within a tpu\_shard\_context, but got unset number\_of\_shards. Assuming 1.'')} in TensorFlow), and documentation (e.g., .md, .td, .pbtxt, .rst, and the markdown type in .ipynb files). (b) Exclude ambiguous assumptions, for example, in the statement: ``\textit{Although these seem generic, they make subtle assumptions about the structure of the graph that is 100\% valid for NNModule graphs but not any graph}'', we only know that there are assumptions regarding the structure of the graph without further details. Such statements were excluded. 

(5) We conducted a formal data extraction according to the data items in Table \ref{tab_dataitem_dl} and Table \ref{tab_dataitem_sca}. In the formal data extraction, the first and third authors independently extracted data from the search results, and the other authors reviewed and discussed the results. The inter-rater agreement on whether to include a piece of data that was measured using the Cohen's Kappa coefficient~\cite{Cohen1960} is 0.946.

(6) Refined the extracted data. As an example, for the content of each SCA (i.e., the ``\textit{Assumption}'' data item), all the search terms were removed (e.g., in TensorFlow, we changed ``\textit{Assume each worker has the same number of replicas}'' to ``\textit{each worker has the same number of replicas}''). This step was done by the first author.

\begin{table}[htbp]
\caption{Extracted data items of each DL framework}
\begin{center}
\begin{tabular}{|p{0.3\columnwidth}|p{0.6\columnwidth}|}
\hline
{\textbf{Data Item}}&{\textbf{Description}}\\
\hline
{Framework} & {Name of the DL framework}\\
\hline
{Release} & {Release version of the DL framework}\\
\hline
{Languages} & {Main programming languages used in the source code}\\
\hline
{Number of Files} & {Number of files of the release}\\
\hline
{Lines of Comments} & {Lines of code comments of the release}\\
\hline
{Lines of Code} & {Lines of code of the release}\\
\hline
{Number of SCAs} & {Number of SCAs included in this study}\\
\hline
{Search Results} & {Number of results from the search}\\
\hline
{Number of Files with SCAs} & {Number of files that have at least one included SCA}\\
\hline
\end{tabular}
\label{tab_dataitem_dl}
\end{center}
\end{table}

\begin{table}[htbp]
\caption{Extracted data items of each assumption}
\begin{center}
\begin{tabular}{|p{0.22\columnwidth}|p{0.68\columnwidth}|}
\hline
{\textbf{Data Item}}&{\textbf{Description}}\\
\hline
{Framework} & {Name of the framework that has SCA \textit{A}}\\
\hline
{File} & {Name of File \textit{F} that has SCA \textit{A}}\\
\hline
{Duplicated File} & {Names of the other files that have SCA \textit{A}}\\
\hline
{Context} & {All related information of SCA \textit{A}, including related artifacts (e.g., source code and decision statements) and impacts of SCA \textit{A}}\\
\hline
{Owner} & {Who made SCA \textit{A}}\\
\hline
{Assumption} & {Content of SCA \textit{A}}\\
\hline
{Constraint} & {Constraints of SCA \textit{A}}\\
\hline
{Complementary Information} & {Additional information that helps to understand SCA \textit{A}}\\
\hline
\end{tabular}
\label{tab_dataitem_sca}
\end{center}
\end{table}

For data analysis, we used descriptive statistics to answer RQ1 (i.e., quantitative data analysis) and Constant Comparison (following the guidelines provided in~\cite{Glaser1967} and~\cite{Adolph2011}) to answer RQ2 and RQ3 (i.e., qualitative data analysis). We did not predefine SCA types and subtypes, but used Constant Comparison to code the SCAs and their context (i.e., incidents/indicators) and generate concepts (i.e., abstraction suggested by the indicators) and categories (i.e., abstraction of the concepts) of the coded SCAs~\cite{Adolph2011}. 
For SCA validity, the extracted data usually has explicit indicators. As an example, considering an SCA from TensorFlow: ``\textit{we must not assume that a path to a file is a path to a parent directory}'', the code is ``\textit{must not assume}'' and the concept/category is ``\textit{invalid}''. If there is no validity indicator, then the code is ``\textit{N/A}'' and the concept/category is ``\textit{valid}''. More examples can be found in Section \ref{RQ2results}.

Regarding SCA content and SCA impacts, the Constant Comparison procedure is much more complex compared to classifying SCAs according to their validity. 
As an example, considering an SCA from TensorFlow: ``\textit{Assume debug\_op\_spec has the format <debug\_op>;<debug\_url>;<gated\_grpc>}'', it was first coded as ``\textit{debug\_op\_spec, format of debug\_op\_spec}''. Since \textit{debug\_op\_spec} is a variable, and by comparison there are other specific variables that concern their format. Therefore, we generated a concept ``\textit{variable format}''. Moreover, in the comparison, we also had other concepts such as ``\textit{variable shape}'' and ``\textit{variable data type}''. Hence we merged these concepts into ``\textit{variable construct}''. Similarly, besides ``\textit{variable construct}'', we also generated others such as ``\textit{variable property and characteristic}'', and ``\textit{variable relationship}''. Therefore, we further generated a ``\textit{variable}'' category. Then we found there was much overlap between the ``\textit{variable}'' category and the ``\textit{tensor}'' category, so we merged the two categories into a new category - ``\textit{tensor and variable}''. Finally, after multiple iterations, since the ``\textit{tensor and variable}'' category cannot be further merged with other categories or concepts, the whole path of the example is ``\textit{tensor and variable format - tensor and variable construct - tensor and variable}''.
Questions such as ``\textit{What is this data a study of?}'' and ``\textit{What is actually happening in the data?}'' were frequently asked to keep the authors theoretically sensitive and transcending during the analysis~\cite{Glaser1967}. We also came up with hundreds of memos to capture our ideas of concepts and categories.

The Constant Comparison steps are: (1) We conducted a pilot study of data analysis with 30 SCAs from TensorFlow and 30 SCAs from PyTorch. The first and third authors independently (a) coded the SCAs (i.e., the codes are words or phrases from the original SCA content), (b) compared the codes to generate concepts, and (c) compared the codes and concepts to generate categories. Each sub-step was iteratively done, i.e., we compared multiple times in each sub-step. The pilot data analysis helped us to (a) reach an agreement on the understanding and usage of the Constant Comparison method and (b) reduce inconsistencies among the authors.
(2) The first and third authors independently analyzed all the data from TensorFlow (1460 SCAs out of 3084, 47.34\%) and PyTorch (569 SCAs out of 3084, 18.45\%), following the same process as mentioned in the pilot data analysis. 
Conflicts were discussed and addressed by all the authors. As an example, one conflict is whether to group processing of specific variables into the ``\textit{tensor and variable}'' category or ``\textit{design}'' category. After discussion, we decided to classify it into the design category, since the ``\textit{tensor and variable}'' category is regarding the tensors and variables themselves instead of using or processing them in certain context.
(3) The first author analyzed the remaining 1055 SCAs (out of 3084, 34.21\%) following the same process used in the pilot data analysis, i.e., (a) coded the SCAs, (b) compared the codes to generate concepts, and (c) compared the codes and concepts to generate categories in multiple iterations. All the other authors reviewed and discussed the results. 

\section{Results} \label{results}
This section only includes results based on the extracted data. We did not add any interpretation from the authors. All the explanation and comparison of the results can be found in Section \ref{interpretation}.
\subsection{Results of RQ1}
Table \ref{tab_overview} shows an overview of the DL frameworks regarding their release, languages, number of files, lines of comments (LOCom), and lines of code (LOC). We identified 3,084 SCAs out of 4,610 search results (66.90\%), which are scattered across 1,775 files (i.e., out of 49,844 files, 3.56\%; one file has 1.74 SCAs on average). This includes 2,278 unique SCAs (out of 3,084, 73.87\%) that have different content and 806 duplicated SCAs (out of 3,084, 26.13\%) that each has the same content as one of the 2,278 SCAs. Only 22 SCAs are duplicated in two DL frameworks. There is no SCA that is duplicated in more than two DL frameworks. Note that an SCA with the same content may have same or different context (e.g., regarding their usage and constraints).
Table \ref{tab_distribution} shows a detailed distribution of SCAs in the DL frameworks. Except for TensorFlow (87.90\%), the other DL frameworks have a final inclusion between 44.44\% and 67.01\% from the search results. TensorFlow has the largest number of identified SCAs in the nine frameworks (1,460 out of 3,084, 47.34\%), while Keras has only 8 SCAs (out of 3,084, 0.26\%). Considering the percentage of files with SCAs (i.e., Number of files with SCAs / Number of total files in a DL framework), Theano is in the first place (6.28\%), followed by PyTorch (4.85\%) and CNTK (4.04\%). On the opposite, PaddlePaddle has only 1.14\% files that include SCAs.

\begin{table}[htbp]
\caption{Overview of the deep learning frameworks}
\begin{center}
\begin{tabular}{|p{0.18\columnwidth}|p{0.12\columnwidth}|p{0.16\columnwidth}|p{0.08\columnwidth}|p{0.12\columnwidth}|p{0.08\columnwidth}|}
\hline
{\textbf{Framework}}&{\textbf{Release}}&{\textbf{Languages}}&{\textbf{Files}}&{\textbf{LOCom}}&{\textbf{LOC}}\\
\hline
{TensorFlow} & {2.3.0} & {Python, C++} & {21,984} & {477K} & {2,527K}\\
\hline
{Keras} & {2.4.0} & {Python} & {243} & {2K} & {21K}\\
\hline
{Caffe} & {1.0} & {C++} & {684} & {13K} & {74K}\\
\hline
{PyTorch} & {1.6.0} & {Python, C++} & {6,943} & {107K} & {1,038K}\\
\hline
{MXNet} & {1.6.0} & {Python, C++} & {3,812} & {95K} & {478K}\\
\hline
{CNTK} & {2.7} & {Python, C++} & {3,096} & {10K} & {179K}\\
\hline
{DL4J} & {1.0.0} & {Java} & {8,641} & {349K} & {948K}\\
\hline
{PaddlePaddle} & {1.8.3} & {Python, C++} & {3,693} & {63K} & {498K}\\
\hline
{Theano} & {1.0.5} & {Python} & {748} & {23K} & {199K}\\
\hline
\end{tabular}
\label{tab_overview}
\end{center}
\end{table}

\begin{table}[htbp]
\caption{Distribution of SCAs}
\begin{center}
\begin{tabular} {|p{0.18\columnwidth}|p{0.08\columnwidth}|p{0.11\columnwidth}|p{0.09\columnwidth}|p{0.09\columnwidth}|p{0.08\columnwidth}|p{0.07\columnwidth}|}
\hline
{\textbf{Framework}}&{\textbf{SCAs}}&{\textbf{Search Results}}&{\textbf{\%}}&{\textbf{Files with SCAs}}&{\textbf{Files}}&{\textbf{\%}}\\
\hline
{TensorFlow} & {1,460} & {1,661} &{87.90\%} & {846} & {21,984} &{3.85\%}\\
\hline
{Keras} & {8} & {18} &{44.44\%} & {6} & {243} &{2.47\%}\\
\hline
{Caffe} & {65} & {97} &{67.01\%} & {25} & {684} &{3.65\%}\\
\hline
{PyTorch} & {569} & {975} &{58.36\%} & {337} & {6,943} &{4.85\%}\\
\hline
{MXNet} & {152} & {270} &{56.30\%} & {103} & {3,812} &{2.70\%}\\
\hline
{CNTK} & {251} & {562} &{44.66\%} & {125} & {3,096} &{4.04\%}\\
\hline
{DL4J} & {420} & {773} &{54.33\%} & {244} & {8,641} &{2.82\%}\\
\hline
{PaddlePaddle} & {61} & {97} &{62.89\%} & {42} & {3,693} &{1.14\%}\\
\hline
{Theano} & {98} & {157} &{62.42\%} & {47} & {748} &{6.28\%}\\
\hline
\textbf{Total} & {3,084} & {4,610} &{66.90\%} & {1,775} & {49,844} &{3.56\%}\\
\hline
\end{tabular}
\label{tab_distribution}
\end{center}
\end{table}



\textbf{Summary}: 3,084 SCAs (including 806 duplicated SCAs) are scattered across 1,775 files in the nine DL frameworks, ranging from 1,460 (TensorFlow) to 8 (Keras) SCAs.

\subsection{Results of RQ2} \label{RQ2results}
Based on the validity of the SCAs, there are four types: Valid SCA, Invalid SCA, Conditional SCA, and Unknown SCA in the DL frameworks. Context is important to understand an assumption. However, due to the page limit, the context of the SCA examples is not provided in the paper, which can be found in the dataset~\cite{Yangdataset2021}.

\textbf{Valid SCA} refers to SCAs that are valid assumptions. This is the most common type in the DL frameworks. 
As an example, in TensorFlow, a developer mentioned: ``\textit{We assume each node has a trivial path to itself so the returned set includes all of the nodes that have ref variables as input or output}''.
\textbf{Invalid SCA} refers to SCAs that are invalid assumptions. It means that stakeholders should not make the assumptions or artifacts (e.g., functions) do not make the assumptions. 
As an example, in MXNet, a developer mentioned: ``\textit{we cannot assume mxnet python bindings are present and in the python path, or that they would point to the specified libmxnet.so}''. 
\textbf{Conditional SCA} refers to SCAs that are either valid or invalid under certain conditions. As an example, in TensorFlow, a developer assumed: ``\textit{an epoch always has something in the buffer until it ends}''. After the description of the SCA, the developer further mentioned: ``\textit{if the input pipeline was slower than the consumers by a lot this might not be true}''. 
\textbf{Unknown SCA} refers to SCAs that their validity is unknown or developers are not sure whether they should make such SCAs. As an example, in PyTorch, a developer assumed: ``\textit{it's unsafe to assume that the passed in pointer in question is the memory pointer in question; it might not be; be sure to read the source code of the Allocator in question to confirm this''}. 

\begin{figure*}[htbp]     
\centering    
\includegraphics[width=15cm,height=9.1cm]{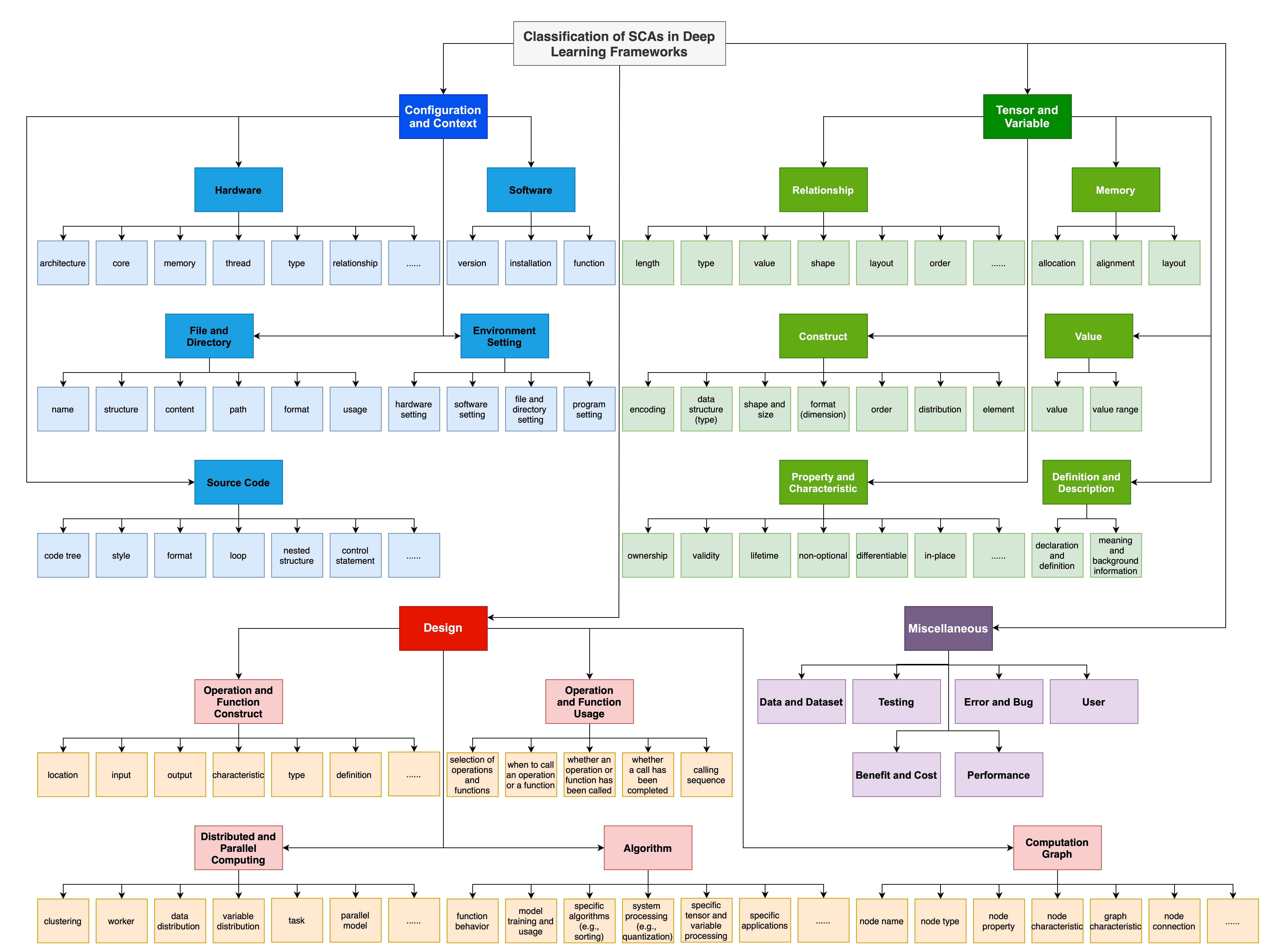}
\caption{A taxonomy of SCAs based on their content}
\label{fig_classification}
\end{figure*}

Based on the content of the SCAs, there are four types: Configuration and Context SCA, Design SCA, Tensor and Variable SCA, and Miscellaneous SCA with a set of subtypes, as shown in Figure \ref{fig_classification}. Because of the page limit, we used ``......'' to represent that a subtype (e.g., the Hardware subtype in the Configuration and Context type) has more items than those presented in the figure. 

\textbf{Configuration and Context SCA} has five subtypes: Hardware SCA, File and Directory SCA, Environment Setting SCA, Software SCA, and Source Code SCA. 
\textbf{Hardware SCA} refers to SCAs concerning various devices (e.g., GPU, CPU, and TPU), such as (1) architecture, core, memory, and thread; (2) type and number; (3) relationship, ownership, and characteristic; (4) strategies of selecting and using devices.
\textbf{File and Directory SCA} is regarding name, structure, content, path, format, and usage of files and directories.
\textbf{Environment Setting SCA} refers to different settings of the system, i.e., (1) hardware setting (e.g., machine's name); (2) software setting (e.g., docker); (3) file and directory setting (e.g., model file); (4) program setting (e.g., compilation and building).
\textbf{Software SCA} is regarding version, installation, and function of certain software (including operation system) in the DL frameworks.
\textbf{Source Code SCA} refers to SCAs concerning the source code of DL frameworks (e.g., code tree, code style, code format, loop, nested structure, data type, control statement, ordinary statement, code optimization, and context of running the code).

\textbf{Tensor and Variable SCA} has six subtypes: Relationship SCA, Property and Characteristic SCA, Construct SCA, Memory SCA, Value SCA, and Definition and Description SCA. Since not all the source code is based on object-oriented programming, we did not explicitly distinguish for example objects from variables. Also we considered matrix, array, tensor, blob, sequence, etc. as the same concept in this work. 
Finally, only the SCAs that explicitly include at least a variable or a tensor were classified into this type, for example, SCAs regarding function/operation inputs or outputs without explicitly referring to variables or tensors were classified into the Design - Operation and Function Construct subtype. 
\textbf{Relationship SCA} refers to SCAs concerning relationships between variables and tensors in various aspects (e.g., length, type, value, shape, and layout).
\textbf{Property and Characteristic SCA} is regarding (1) ownership, inheritance, validity, and lifetime of variables and tensors; (2) existence of a property; (3) whether a variable or tensor is non-optional, differentiable, in-place, printable, immutable, vectorizable, discardable, broadcastable, trainable, stateful, or executable; (4) others such as whether a variable is a SparseVariable, safe to copy, or has duplication.
\textbf{Construct SCA} includes aspects of encoding, data structure (type), shape and size, format (dimension), order, distribution, and element (e.g., number of specific elements and existence of an element in a variable).
\textbf{Memory SCA} is regarding memory allocation, alignment, and layout of variables and tensors.
\textbf{Value SCA} refers to the exact value or value range of variables.
\textbf{Definition and Description SCA} includes SCAs regarding (1) declaration and definition and (2) meaning and background information of variables and tensors.

\textbf{Design SCA} has five subtypes: Operation and Function Construct SCA, Operation and Function Usage SCA, Algorithm SCA, Computation Graph SCA, and Distributed and Parallel Computing SCA. 
\textbf{Operation and Function Construct SCA} includes location, construct of input and output, characteristic, type, definition, and example (help for understanding) of operations and functions.
\textbf{Operation and Function Usage SCA} includes (1) selection of operations and functions; (2) when to call an operation or a function; (3) whether an operation or a function has been called; (4) whether a call has been completed; (5) calling sequence of operations and functions.
\textbf{Algorithm SCA} is the most common case in the Design type, which includes many aspects, such as 
(1) function behavior; 
(2) algorithms of training and using models (e.g., model input and output, model layer, model signature, learning rate, gradient, loss, dropout, weight, activation, (un)pooling, and normalization); 
(3) specific algorithms like rounding, sorting, Fast Fourier Transform, Sigmoid, Softmax, Recurrent Neural Network, and reinforce learning; 
(4) system processing such as broadcasting, computation, concatenation, quantization, transformation, tracing, resource management, thread management, and memory strategies (e.g., memory sharing, mapping, and layout strategies); 
(5) processing of specific variables, tensors, data, file, etc.;
(6) specific applications like bounding boxes in object detection, pixels in image processing, and region of interest; 
(7) others like usage of design patterns. 
\textbf{Computation Graph SCA} refers to SCAs concerning the static or dynamic computation graph (e.g., TensorFlow) or layer-based graph (e.g., Caffe) in the DL frameworks, including node (e.g., name, type, property, characteristic, data type, life span, connection, and dependency), graph layer, control flow, context, and graph characteristic (e.g., validity, loopless, and variance). 
\textbf{Distributed and Parallel Computing SCA} is regarding SCAs of the distributed and parallel computing in DL frameworks, such as cluster, worker, data distribution, variable distribution, task, and parallel model.

\textbf{Miscellaneous SCA} has six subtypes: Data and Dataset SCA, Testing SCA, Error and Bug SCA, User SCA, Benefit and Cost SCA, and Performance SCA.
\textbf{Data and Dataset SCA} is regarding the data or dataset used in the DL frameworks. 
\textbf{Testing SCA} is regarding testing in the DL frameworks. There are many SCAs in the testing files/functions (with a prefix or suffix of ``\textit{test}''). However, unless they are explicitly regarding testing, we did not classify them into this subtype.
\textbf{Error and Bug SCA} includes (1) existence of errors and bugs; (2) reasons of errors and bugs; (3) how to treat errors and bugs. \textbf{User SCA} includes SCAs of user behavior and knowledge. 
\textbf{Benefit and Cost SCA} refers to SCAs regarding benefits or costs of something in the system.
\textbf{Performance SCA} concerns system performance in the DL frameworks.

\textbf{Summary}: There are four types of validity of SCAs: Valid SCA, Invalid SCA, Conditional SCA, and Unknown SCA, and four types of SCAs based on their content: Configuration and Context SCA, Design SCA, Tensor and Variable SCA, and Miscellaneous SCA.

\subsection{Results of RQ3}

In the identified SCAs, both valid and invalid SCAs can have impacts on their involved DL frameworks, containing three impact patterns: ``what things would be'', ``what we can/need to do'', and ``negative impact''. ``what things would be'' refers to impacts to one or more artifacts (e.g., variables) regarding the value, shape, state, etc. by an SCA. 
An example in MXNet is: ``\textit{Assume the input has size *k* on axis 1, then both ``gamma'' and ``beta'' have shape *(k,)*.}'' In the example, the shape of ``\textit{gamma}'' and ``\textit{beta}'' can be inferred based on the SCA.
``what we can/need to do'' refers to things that can/need to be done if there is an SCA. 
An example in TensorFlow is: ``\textit{FileExists cannot differentiate between existence of a file or a directory, hence we need an additional test as we must not assume that a path to a file is a path to a parent directory.}'' The example shows that based on the design of ``\textit{FileExists}'' (a function) and the invalid SCA, there is a need to conduct an additional test when calling the ``\textit{FileExists}'' function.
``negative impact'' refers to negative impacts (e.g., system vulnerabilities and failures) caused by an SCA. 
For example, in PyTorch, there is a function ``\textit{RemoveInplaceOps}'', which is used to remove all in-place operations and replace them with out-of-place equivalents. In the code comment of the function, the developer mentioned: ``\textit{NOTE: this is NOT SAFE, since it assumes that the LHS is not aliased by another value.}'' The example shows that though the SCA itself is valid, the SCA leads to that running the function is unsafe.
Also certain technical debt is induced when making SCAs. An example in TensorFlow is: ``\textit{TODO: Looks like there is an assumption that weight has only one user. We should add a check here.}''


SCAs are not independent in software development. The content of the SCAs includes many types of software artifacts, such as directory, file, module, function, and variable, and therefore, these artifacts are naturally related to the SCAs. 
Besides the artifacts mentioned in the SCA content, there are source code written based on certain SCAs. An example in PaddlePaddle is: ``\textit{Random outputs give accuracy about 0.33, we assume valid accuracy to be at least 0.5.}'' Based on this SCA, the related source code is: ``\textit{assert fp32\_acc \textgreater 0.5;} \textit{assert int8\_acc \textgreater 0.5;}''. 
Moreover, many decisions are made based on certain SCAs, i.e., design decisions and SCA decisions. Design decisions refer to the decisions of detailed design, such as module design, function design, and variable design. An example in a function of MXNet is: ``\textit{here we calculate the output number (exclude min/max, in order to calculate min/max index from mirror node) based on assumption that there is only 1min and 1max output from mirror node (which is currently true).}''
SCA decisions refer to the decisions regarding SCAs themselves. An example in PaddlePaddle is: ``\textit{Revisit this assumption. what if \#blocks \textgreater \#pservers. NOTE: assume blocks of the same variable is not distributed on the same pserver, only change param/grad varnames for trainers to fetch.}'' The example shows that the developers would revisit this SCA in the DL framework.


Though SCAs are important in software development, the impacts of SCAs usually have a scope. There are two types of scope: SCA Owner and SCA Constraint. 
All the SCAs were made by certain developers. Terms in the extracted data include for example ``\textit{we assume}'', ``\textit{let's assume}'', and ``\textit{my assumption}''. However, many SCAs were also explicitly mapped to certain artifacts, which forms the SCA Owner type. SCA Owner refers to artifacts (e.g., a specific module, function, or variable) that made the SCA, which can limit the impact scope. 
As an example, in Theano, there is an SCA: ``\textit{If ``outputs\_info'' is an empty list or None, ``scan'' assumes that no tap is used for any of the outputs.}'', while ``\textit{scan}'' is a function defined in ``\textit{scan.py}''. Therefore, the impacts of this SCA are limited in the ``\textit{scan}'' function. 
Moreover, there are SCAs that use ``\textit{this function assumes}'', ``\textit{the algorithm assumes}'', ``\textit{the caller assumes}'', ``\textit{the class assumes}'', ``\textit{the implementation assumes}'', ``\textit{the method assumes}'', ``\textit{the op assumes}'', etc. Such terms also indicate the scope of SCA impacts. 
Some SCAs have a broader impact scope, using terms such as ``\textit{a lot of code assumes}'', ``\textit{all builtin operators assume}'', ``\textit{all of the tests assume}'', ``\textit{all other functions called from RunAllTests() assume}'', ``\textit{all sorts of things assume}'', and ``\textit{applications assume}''. 
Finally, certain SCAs limit their scope to a DL framework. As an example, in TensorFlow, there is an SCA: ``\textit{Keras assumes that batch dimension is the first dimension for Batch Normalization.}''
Another type of the scope of SCA impacts is SCA Constraint, which refers to conditions that an SCA should satisfy. The most common terms used in the identified SCA constraints are ``\textit{if}'' and ``\textit{when}''. An example in PyTorch is: ``\textit{If :math:`y = 1` then it assumed the first input should be ranked higher (have a larger value) than the second input.}''
The example shows that if the constraint of the SCA is not satisfied, the SCA would not exist when running the system, and therefore, there would be no impacts of the SCA to the running system. 
There are also other terms: ``\textit{unless}'', ``\textit{after}'', ``\textit{as long as}'', ``\textit{by default}'', ``\textit{except for}'', ``\textit{for (certain things)}'', ``\textit{in (certain things)}'', ``\textit{while}'', and ``\textit{with}''. 

\textbf{Summary}: Both valid and invalid SCAs may have an impact within a specific scope (e.g., in a function) on the DL frameworks. Certain technical debt is induced when making SCAs. There are source code written and decisions made based on SCAs.

\section{Discussion} \label{discussion}
\subsection{Interpretation of Results} \label{interpretation}
This section provide the explanation and comparison of the results.

\textbf{RQ1: The distribution of SCAs in DL frameworks}

Every selected framework has SCAs, which indicates that SCAs are common in DL frameworks. TensorFlow has the largest number of the identified SCAs in the nine frameworks. One obvious reason is that TensorFlow is the largest project with the most number of files, lines of code, and lines of comments, compared to the other eight DL frameworks. On the opposite, Keras has 8 SCAs. The reason is that Keras is a library (interface) built based on backends (e.g., TensorFlow), which has only 21K lines of code. 
Regarding the number of total files that have SCAs, Theano is in the first place, while PaddlePaddle has only 1.14\% files that include SCAs. There are four possible reasons: (1) not all of the files include code comments (e.g., documentation files and binary files); (2) only the latest version of the nine DL frameworks was used to extract SCAs. Since SCAs have a dynamic nature (i.e., can evolve over time), certain SCAs may be removed in previous versions of the DL frameworks; (3) assumptions may stay implicit and undocumented during the development, which are not SCAs; and (4) certain SCAs are unimportant or obvious and there is no need to document them in the code comments. 

Similar to duplicated code, the duplicated SCAs can also increase system vulnerability and decrease system maintainability. The potential reasons are that: (1) multiple places need to be updated when such SCAs change; (2) if an SCA turns out to be invalid, the duplicated ones may also turn out to be invalid when they have same (or similar) context; and (3) if an SCA and its duplicates have different context (e.g., constraints, decisions, and usage) without proper documentation, maintaining these SCAs in the same way may lead to invalid SCAs. It is even worse if duplicated SCAs are described in different sentences. For example, in TensorFlow, there are two SCAs: (1) ``\textit{This function assumes the both operands are verified to have value attributes of broadcastable types.}'' and (2) ``\textit{This function assumes the two operands are verified to have value attributes of broadcastable types.}'' If we use one of the SCAs to search for the other one, we would not get any result, since the first one uses ``\textit{both operands}'' while the second one uses ``\textit{two operands}''.
Furthermore, only 22 SCAs are duplicated in two DL frameworks, and there is no SCA shared by at least three DL frameworks. There are two reasons: (1) stakeholders in different projects may make various SCAs or describe the same SCA differently; (2) the source code in the DL frameworks are not widely reused.


\textbf{RQ2: The types of SCAs in DL frameworks}

There are three dimensions for assumption classification in software development, i.e., explicitness and documentation, validity, and content of assumptions. From the explicitness and documentation perspective, Roeller \textit{et al.}~\cite{Roeller2006} proposed a classification of architectural assumptions, i.e., (1) Implicit and Undocumented Assumption, (2) Explicit and Undocumented Assumption, (3) Explicit and Explicitly Undocumented Assumption, and (4) Explicit and Documented Assumption. The SCAs in this study belong to the last type, i.e., Explicit and Documented Assumption. 

From the validity perspective, assumptions are normally classified as Valid Assumption and Invalid Assumption. In addition, Fassi \textit{et al.}~\cite{Fassi2020} added another type ``Awaiting Evaluation'', and in our earlier work~\cite{Yangtool2021}, we came up with a similar type (called Unknown Assumption). In this work, besides Valid, Invalid, and Unknown (Awaiting Evaluation) Assumption, we further identified a new type: Conditional Assumption, i.e., assumptions can be valid or invalid under certain conditions. This is consistent with the context dependent characteristic of assumptions in development. 

From the content perspective, in our earlier work~\cite{Yangsms2018}, we classified assumptions in software development into five types according to the SWEBOK~\cite{Bourque2014}, i.e., Requirements Engineering Assumption, Design Assumption, Construction Assumption, Testing Assumption, and Maintenance and Evolution Assumption. Other related studies focus on a specific type of assumptions (e.g., Architectural Assumption). For example, Lago and van Vliet~\cite{Lago2005} classified architectural assumptions into three types: Technical Assumption, Organizational Assumption, and Management Assumption. The existing classifications are either too general or focusing on a specific part of development (e.g., requirements engineering or architecture design), which is not suitable for DL frameworks. Therefore, in this work, we used Constant Comparison to come up with a taxonomy composed of four types: Configuration and Context SCA, Tensor and Variable SCA, Design SCA, and Miscellaneous SCA.

\textbf{RQ3: The impacts of SCAs in DL frameworks}

In this work, we identified impacts from not only invalid SCAs but also valid ones. Besides the negative impacts (e.g., system failure) already discussed in literature, we also found certain technical debt that is induced when making SCAs. The nature of SCAs is uncertainty and such uncertainty usually cannot be removed when writing the code, leading to the need of managing the SCAs later in the project, which can incur technical debt. Furthermore, there are other impact patterns: ``what things would be'' and ``what we can/need to do'', which are also not mentioned in previous studies (e.g., \cite{Yangsms2018} and \cite{Eliasson2014}). The reasons are that: (1) the research scope of the previous studies is set to invalid (or not well-managed) assumptions; (2) the number of assumptions collected in the previous studies is limited. 

Since the SCAs were identified from code comments instead of project documents, most of the related artifacts are source code artifacts (e.g., directories, files, modules, functions, and variables). The results also indicate that there are source code written and decisions (i.e., design decisions and SCAs decisions) made based on SCAs. This is consistent with previous studies (e.g., \cite{Yangsms2018} and \cite{Eliasson2014}) on assumptions and their management in software development. 

As mentioned in the results, the impacts of SCAs usually have a scope. One reason is that the SCAs are mostly related to code artifacts (e.g., modules, functions, and variables). Another reason is that there are SCAs triggered by constraints. If the constraints of an SCA are not met, the SCA would not exist when running the system, and therefore, there would be no impacts of the SCA to the running system. Though SCAs may have an impact scope, they are still important in development. For example, considering an invalid SCA that causes only a function to fail, such failure however may spread and lead to a series of errors among dependent functions and then eventually result in system crashes.

\subsection{Implications for Researchers}
Considering the provided SCA examples, if we remove the search terms, there is no other indicator to identify whether the sentences are assumptions, i.e., the linguistic patterns of implicit assumptions remain unknown. This is the major reason of impeding researchers from identifying implicit but documented assumptions in DL framework projects as well as other types of open source software projects. Therefore, there is a need to extract implicit assumptions from/with stakeholders who made the assumptions and then identify and report the linguistic patterns of such assumptions.

This is the first work regarding SCAs in DL frameworks. Though we included nine popular DL frameworks, there is still a need to investigate SCAs in more frameworks, as well as DL applications that are built based on the frameworks. The data of this study was extracted from code comments, while mining SCAs from other sources, such as issues on GitHub and discussions on Stack Overflow, can provide more valuable and comprehensive data for further analysis and comparison.

During development, a valid SCA can turn out to be invalid or vice versa, or an SCA can transform to another type of artifact or vice versa. The distribution, explicitness, validity, content, and impacts of SCAs can all evolve over time. Therefore, we encourage researchers to investigate the evolution of SCAs in DL frameworks.

Not all of the SCAs are important (or negative) to the development or usage of DL frameworks. For example, some SCAs identified in this study are only used in examples for better understanding the functions of DL frameworks. Such SCAs have limited impacts on running the code. Therefore, there is a need to study how to identify important SCAs, which can also help practitioners to better deal with SCAs in their projects.

The relationship between SCAs and technical debt has not been studied in literature, such as what SCAs are included in the description of technical debt or how SCAs incur technical debt. Considering the importance of managing assumptions and technical debt in software projects, there is a need to investigate such relationship in software development.

\subsection{Implications for Practitioners}
Making assumptions is one way to deal with uncertainties in development. It is a good practice to make assumptions explicit and documented. The SCAs in the DL frameworks are usually described in one or several sentences. This leads to a problem that SCAs may lack information (e.g., constraints and related artifacts) for further analysis. We encourage developers to follow certain templates to record SCAs in a structure way for a better understanding of them, when the documentation effort is acceptable.

Though SCAs are not all negative in development, it is encouraged to confirm SCAs if their uncertainty can be removed, transforming the SCAs to other types of artifacts (e.g., decisions and source code) and consequently reducing SCAs in projects. For those important SCAs, it is recommended to add checks if possible, which can detect violations of SCAs and throw warnings or exceptions when running the code of DL frameworks. 

Similar to duplicated code that leads to code clones, it is not a good practice to have duplicated SCAs in multiple places in the system, since they will increase system vulnerability and decrease system maintainability.

Though not all the identified SCAs are important to the users of the DL frameworks, being aware of the SCAs can help the users to better understand the rationale behind the frameworks, and further increase the quality of the DL applications developed based on the frameworks. Specifically, we encourage that the users should pay enough attention to invalid, unknown, and conditional SCAs, as well as what need to be done to deal with the SCAs. This is especially important if the users need to customize functions of the frameworks in their projects. Framework developers are also encouraged to provide specific tools or plugins that can remind framework users of the SCAs.

\section{Threats to Validity} \label{threatstovalidity}
The threats to the validity of the study can be divided in four types, i.e., construct validity, external validity, internal validity, and reliability~\cite{Runeson2012}. In this study, we excluded internal validity because we did not investigate any causal relationships.

\textbf{Construct validity}.
One threat is whether the collected data can be properly used to answer the RQs. To reduce the threat, we first designed a protocol of the study according to our experience and literature review; all the authors reviewed the protocol iteratively, and reached an agreement on the design (e.g., the data items and the mapping between the RQs and data items).
In this work, we depended on human activities, which would induce personal bias. To reduce the threat, we conducted a two-round pilot study in data extraction, and reached a high agreement on both the second round pilot (Cohen's Kappa coefficient of 0.950) and formal (Cohen's Kappa coefficient of 0.946) data extraction. For qualitative data analysis, we followed a systematic Constant Comparison process. By constantly trying to fit words to represent the patterns, we found the best fit names to build the concepts and categories, and therefore, we mitigated the threat. We also conducted a pilot study, which can reduce the threat of personal bias. 
Moreover, regarding the situation that stakeholders may not use the ``\textit{assumption}'' term to express an assumption, such assumptions are not explicit, and therefore, they are not SCAs, which is out of the scope of this study.  

\textbf{External validity}.
Based on the related work, we included nine of the most popular DL frameworks in this work. Therefore, the results are highly representative in DL framework development.

\textbf{Reliability}.
To improve reliability, (1) we conducted a pilot data extraction and a pilot data analysis to reach a consensus and reduce inconsistencies between the researchers; (2) the first and third authors independently extracted and analyzed the data and the other authors reviewed the results; (3) all the outputs of each iteration were documented; and (4) we provided the dataset constructed in this study online~\cite{Yangdataset2021} for other researchers to replicate this study.

\section{Conclusions} \label{conclusions}
In this study, we extracted and analyzed SCAs from nine popular DL framework projects on GitHub. The results are that: (1) 3,084 SCAs are scattered across 1,775 files in the nine DL frameworks, ranging from 1,460 (TensorFlow) to 8 (Keras) SCAs. (2) There are four types of validity of SCAs: Valid SCA, Invalid SCA, Conditional SCA, and Unknown SCA, and four types of SCAs based on their content: Configuration and Context SCA, Design SCA, Tensor and Variable SCA, and Miscellaneous SCA. (3) Both valid and invalid SCAs may have an impact within a specific scope (e.g., in a function) on the DL frameworks. Certain technical debt is induced when making SCAs. There are source code written and decisions made based on SCAs.
This is the first work on investigating SCAs in DL frameworks, which helps researchers and practitioners to have a comprehensive understanding on this topic. We also provide the first SCA dataset~\cite{Yangdataset2021} for the community when conducting further research and practice in this area (e.g., automatic assumption identification and impact analysis of SCAs). 

For future work, we aim to (1) mine SCAs from diverse dimensions (e.g., issue reports) and sources (e.g., Stack Overflow and DL applications) to further understand and analyze SCAs in DL frameworks, (2) investigate how developers manage SCAs in DL frameworks, i.e., the evolution of SCAs in DL frameworks (e.g., whether SCAs are revised or removed over time) as well as its impacts to system evolution (e.g., how SCAs would induce technical debt or defects), and (3) provide tools to help DL framework developers and users to be aware of and better understand the SCAs in DL framework and application development.

\begin{acks}
This work has been partially supported by the National Key R\&D Program of China with Grant No. 2018YFB1402800.
\end{acks}

\balance

\bibliographystyle{acm-reference-format}
\bibliography{ref}
\end{document}